\documentclass[12pt]{article}
\usepackage{axodraw,a4wide}

\newcommand{\ie}{i.e.\ }
\newcommand{\eg}{e.g.\ }
\newcommand{\FA}{{\sl FeynArts}}
\newcommand{\mma}{{\sl Mathematica}}
\newcommand{\FC}{{\sl FormCalc}}
\newcommand{\FO}{{\sl FORM}}
\newcommand{\LT}{{\sl LoopTools}}
\newcommand{\MW}{M_{\rm W}}
\newcommand{\MZ}{M_{\rm Z}}

\newcommand{\cpc}[3]{{\sl Comp. Phys. Commun.} {\bf #1} (19#2) #3}
\newcommand{\fp}[3]{{\sl Fortschr. Phys.} {\bf #1} (19#2) #3}
\newcommand{\np}[3]{{\sl Nucl. Phys.} {\bf #1} (19#2)~#3}

\newcommand{\pr}[3]{{\sl Phys. Rev.} {\bf #1} (19#2) #3}
\newcommand{\zp}[3]{{\sl Z. Phys.} {\bf #1} (19#2) #3}

\begin{document}

\thispagestyle{empty}

\hfill KA--TP--5--1999

\hfill hep-ph/9905354

\vspace{3cm}

\begin{center}
\begin{Large}
Generating and Calculating One-loop Feynman Diagrams

with \FA, \FC, and \LT

\end{Large}
\end{center}

\vspace{3cm}

\begin{center}
T.~Hahn\\%
Institut f\"ur Theoretische Physik, Universit\"at Karlsruhe\\%
D--76128 Karlsruhe, Germany
\end{center}

\vfill

Abstract:

A set of programs is presented for automatically generating and
calculating Feynman diagrams. Diagrams are generated with \FA, then
algebraically simplified using a combination of \mma\ and \FO\ implemented
in the package \FC, and finally evaluated numerically using the \LT\
package. \FC\ works either in dimensional regularization or in constrained
differential renormalization, the latter of which is equivalent at the
one-loop level to regularization by dimensional reduction. \FC\ combines
the speed of \FO\ with the powerful instruction set of \mma, and the
latter greatly eases further processing of the results (\eg selecting or
modifying terms). The output is in a form well suited for numerical
evaluation, which is then straightforward using the implementations of the
one-loop integrals in \LT.

\setcounter{page}{0}
\clearpage

%------------------------------------------------------------------------

\section{Introduction}

Explaining the necessity of one-loop calculations in the light of
modern-day colliders is like carrying owls to Athens; and as there is no
lack of motivation, several program packages have been developed to aid
these calculations. However, these programs generally tackle only part of
the problem, so there is still considerable work left in making these
programs work together.

In this paper three packages, \FA, \FC, and \LT, are presented which
work hand in hand. The user has to supply only small driver programs whose
main purpose is to specify the necessary input parameters. This makes the
whole system very ``open'' in the sense that the results are returned as
\mma\ expressions which can easily be manipulated, \eg to select or modify
terms.

Since one-loop calculations can range anywhere from a handful to several
hundreds of diagrams (particularly so in models with many particles like
the MSSM), speed is an issue, too. \FC, the program which does the
algebraic simplification, therefore uses \FO\ \cite{Ve91} for the
time-consuming parts of the calculation. Owing to \FO's speed, \FC\ can
process, for example, the 1000-odd one-loop diagrams of W--W scattering in
the Standard Model \cite{DeH98} in about 5 minutes on an ordinary Pentium
PC.

The following table summarizes the steps in a one-loop calculation and
the distribution of tasks among the programs \FA, \FC, and \LT\/:
\begin{center}
\begin{tabular}{|c|c|l|l}
\cline{1-1} \cline{3-3}
		&& $\bullet$ Create the topologies \\
Diagram		&& $\bullet$ Insert fields \\
generation	&& $\bullet$ Apply the Feynman rules \\
		&& $\bullet$ Paint the diagrams
& \smash{\raise 4.3ex%
  \hbox{$\left.\vrule width 0pt depth 5ex height 0pt\right\}$ \FA}} \\
\cline{1-1} \cline{3-3}
\multicolumn{1}{c}{$\downarrow$} \\
\cline{1-1} \cline{3-3}
		&& $\bullet$ Contract indices \\
Algebraic	&& $\bullet$ Calculate traces \\
simplification	&& $\bullet$ Reduce tensor integrals \\
		&& $\bullet$ Introduce abbreviations
& \smash{\lower 1.1ex%
  \hbox{$\left.\vrule width 0pt depth 10.5ex height 0pt\right\}$ \FC}} \\
\cline{1-1} \cline{3-3}
\multicolumn{1}{c}{$\downarrow$} \\
\cline{1-1} \cline{3-3}
		&& $\bullet$ Convert \mma\ output \\
Numerical	&& \qquad to Fortran code \\
evaluation	&& $\bullet$ Supply a driver program \\
		&& $\bullet$ Implementation of the integrals
& $\left.\mathstrut\right\}\,$ \LT \\
\cline{1-1} \cline{3-3}
\end{tabular}
\end{center}
The following sections describe the main functions of each program.
Also, to demonstrate how the programs are used together, the \FC\ package
contains two sample calculations in the electroweak Standard Model:
$ZZ\to ZZ$ \cite{DeDH97} and $e^+e^-\to\bar t\,t$ \cite{BeMH91}.

\section{\FA}

\FA\ is a \mma\ package for the generation and visualization of Feynman
diagrams and amplitudes \cite{KuBD91}. It works in three basic steps
sketched in the following diagram:
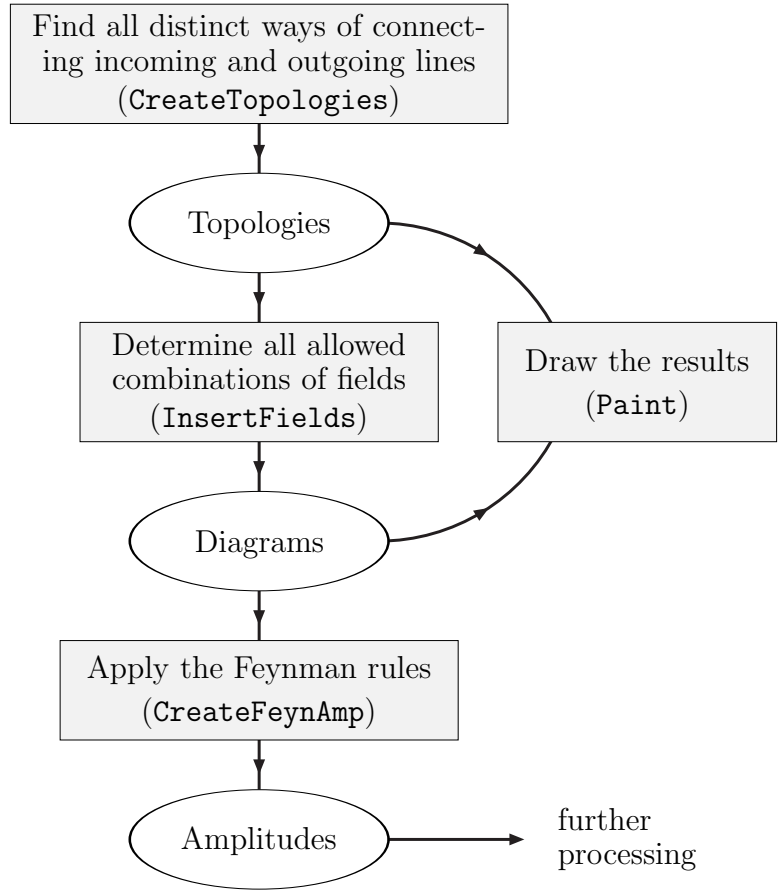
\begin{center}
\unitlength=1bp%
\begin{picture}(310,342)(20,15)
\SetScale{.75}
\SetWidth{1.5}
\ArrowLine(150,411)(150,380)
\ArrowLine(150,335)(150,308)
\ArrowLine(150,251)(150,222)
\ArrowLine(150,175)(150,148)
\ArrowLine(150,101)(150,72)

\Line(200,50)(280,50)
\ArrowLine(279,50)(280,50)
\Text(225,41)[lb]{further}
\Text(225,36)[lt]{processing}

\ArrowArcn(210,260)(100,90,27)
\ArrowArc(210,300)(100,-90,-27)
\SetWidth{.5}

\GBox(25,410)(275,470){.95}
\Text(112.5,338)[b]{Find all distinct ways of connect-}
\Text(112.5,325)[b]{ing incoming and outgoing lines}
\Text(112.5,322)[t]{({\tt CreateTopologies})}
\GOval(150,360)(25,65)(0){1}
\Text(112.5,269)[]{Topologies}
\GBox(60,250)(240,310){.95}
\Text(112.5,218)[b]{Determine all allowed\vphantom{p}}
\Text(112.5,205)[b]{combinations of fields\vphantom{p}}
\Text(112.5,202)[t]{({\tt InsertFields})}
\GBox(270,250)(410,310){.95}
\Text(255,212)[b]{Draw the results\vphantom{p}}
\Text(255,207)[t]{({\tt Paint})}
\GOval(150,200)(25,65)(0){1}
\Text(112.5,149)[]{Diagrams}
\GBox(50,100)(250,150){.95}
\Text(112.5,96)[b]{Apply the Feynman rules}
\Text(112.5,91)[t]{({\tt CreateFeynAmp})}
\GOval(150,50)(25,65)(0){1}
\Text(112.5,37)[]{Amplitudes}
\end{picture}
\end{center}

The first step is to create all different topologies for a given number of
loops and external legs. For example, to create all one-loop topologies
for a $2\to 2$ process except those containing a tadpole, the following
call to {\tt CreateTopologies} is used:
\begin{verbatim}
   tops = CreateTopologies[ 1, 2 -> 2, ExcludeTopologies -> {Tadpoles} ]
\end{verbatim}

In the second step, the actual particles in the model have to be
distributed over the topologies in all allowed ways. E.g.\ the diagrams
for $e^+e^-\to \gamma\gamma$ are produced with
\begin{verbatim}
   inss = InsertFields[ tops, {-F[2, {1}], F[2, {1}]} -> {V[1], V[1]} ]
\end{verbatim}
where \verb=F[2, {1}]= is the electron, \verb=-F[2, {1}]= its
antiparticle, the positron, and \verb=V[1]= the photon. The fields and
their couplings are defined in a special file, the model file, which the
user can supply or modify. Model files are currently provided for QED, the
electroweak Standard Model, and QCD; a MSSM model file is in preparation.

The diagrams may be painted with \verb=Paint[inss]=. Finally, the
analytic expressions for the diagrams are obtained by
\begin{verbatim}
   amps = CreateFeynAmp[ inss ]
\end{verbatim}

\section{\FC}

The evaluation of the \FA\ output proceeds in two steps:
\begin{enumerate}
\item
The symbolic expressions for the diagrams are simplified algebraically
with \FC\ which returns the results in a form well suited for numerical
evaluation.
\item
The \mma\ expressions then need to be translated into a Fortran program.
(The numerical evaluation could, in principle, be done in \mma\ directly,
but this becomes very slow for large amplitudes.) The translation is done
by {\sl NumPrep} which is part of the \FC\ package. For compiling the
generated code one needs a driver program ({\tt num.F}, also in \FC), and
the numerical implementations of the one-loop integrals in \LT.
\end{enumerate}

The structure of \FC\ is simple: it prepares the symbolic expressions of
the diagrams in an input file for \FO, runs \FO, and retrieves the
results. This interaction is transparent to the user. \FC\ combines the
speed of \FO\ with the powerful instruction set of \mma\ and the latter
greatly facilitates further processing of the results. The following
diagram shows schematically how \FC\ interacts with \FO\/: 
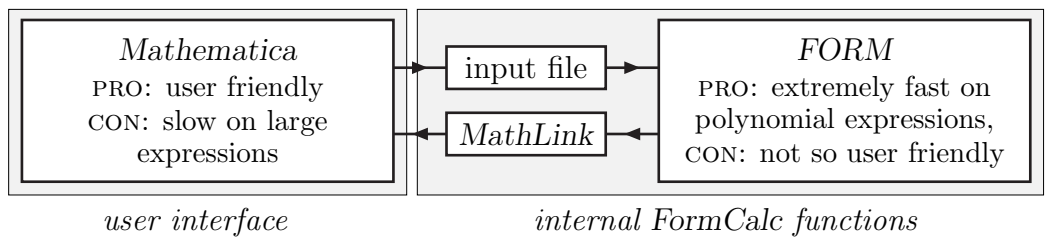
\begin{center}
\begin{picture}(390,87)(-40,9)
\SetWidth{0}
\GBox(-40,25)(110,95){.95}
\GBox(114,25)(350,95){.95}
\SetWidth{1}
\BBox(-35,30)(105,91)
\BBox(205,30)(345,91)
%\SetWidth{1.5}
%\ArrowLine(105,72)(205,72)
\BBox(125,65)(185,81)
\ArrowLine(185,73)(205,73)
\ArrowLine(105,73)(125,73)
\Text(155,72)[]{input file}
\BBox(125,40)(185,56)
\ArrowLine(205,48)(185,48)
\ArrowLine(125,48)(105,48)
\Text(155,48)[]{\sl MathLink}
\Text(35,80)[]{\mma}
\Text(35,65)[]{\small {\sc pro:} user friendly}
\Text(35,52)[]{\small {\sc con:} slow on large}
\Text(35,40)[]{\small expressions}
\Text(275,80)[]{\FO}
\Text(275,65)[]{\small {\sc pro:} extremely fast on}
\Text(275,53)[]{\small polynomial expressions,}
\Text(275,40)[]{\small {\sc con:} not so user friendly}
\Text(30,20)[t]{\it user interface}
\Text(230,20)[t]{\it internal \FC\ functions}
\end{picture}
\end{center}
\FC\ can work either in dimensional regularization or in constrained
differential renormalization (CDR) \cite{techniques}, the latter of which
is equivalent at the one-loop level to regularization by dimensional
reduction \cite{HaP98}.

The main function in \FC\ is {\tt OneLoop} (the name is not strictly
correct since it works also with tree graphs). It is used like this:
\begin{verbatim}
   << FormCalc.m;
   $Dimension = D;
   amps = << myamps.m;
   alldiags = OneLoop[amps]
\end{verbatim}
The file {\tt myamps.m} is assumed here to contain amplitudes generated
by \FA. The dimension---{\tt D} for dimensional regularization or {\tt 4}
for dimensional reduction / CDR---is set with {\tt \$Dimension}. 
Note that {\tt OneLoop} needs no declarations of the kinematics of the
underlying process; it uses the information \FA\ hands down.

Even more comprehensive than {\tt OneLoop}, the function {\tt ProcessFile}
can process entire files. It collects the diagrams into blocks such that
index summations (\eg over fermion generations) can later be carried out
easily, \ie only diagrams which are summed over the same indices are put
in one block. {\tt ProcessFile} is invoked \eg as
\begin{verbatim}
   ProcessFile["vertex.amp", "results/vertex"]
\end{verbatim}
which reads the \FA\ amplitudes from {\tt vertex.amp} and produces
files of the form {\tt results/vertex{\it id}.m}, where {\it id} is an
identifier for a particular block.

{\tt OneLoop} and {\tt ProcessFile} return expressions where spinor
chains, dot products of vectors, and Levi-Civita tensors contracted with
vectors have been collected and abbreviated. A term in such an expression
may look like
\begin{verbatim}
   C0i[cc1, MW2, S, MW2, MZ2, MW2, MW2] *
     (P12*S*(-8*a2*MW2 + 4*a2*MW2*S2 - 28*a2*CW^2*MW2*S2 +
             16*a2*CW^2*S*S2 + 4*a2*C2*MW2*SW^2) +
      O47*S*(-32*a2*CW^2*MW2*S2 + 8*a2*CW^2*S2*T + 8*a2*CW^2*S2*U) -
      P13*S*(-64*a2*CW^2*MW2*S2 + 16*a2*CW^2*S2*T + 16*a2*CW^2*S2*U))
\end{verbatim}
The first line stands for the tensor coefficient function
$C_1(\MW^2, s, \MW^2, \MZ^2, \MW^2, \MW^2)$ which is multiplied with a
linear combination of abbreviations like {\tt O47} or {\tt P12} with
certain coefficients. These coefficients contain the Mandelstam variables
{\tt S}, {\tt T}, and {\tt U} and some short-hands for parameters of the
Standard Model, \eg ${\tt a2} = \alpha^2$.

The abbreviations like {\tt O47} or {\tt P12} are introduced automatically
and can significantly reduce the size of an amplitude. The definitions of
the abbreviations can be retrieved by {\tt Abbreviations[]} which returns
a list of rules such that \verb=result //. Abbreviations[]= gives the
full, unabbreviated expression.

\section{\LT}

\LT\ supplies the actual numerical implementations of the one-loop
functions needed for programs made from the \FC\ output. It is based on
the reliable package {\sl FF} \cite{vOV90} and provides in addition to the
scalar integrals of {\sl FF} also the tensor coefficients in the
conventions of \cite{De93}. \LT\ offers three interfaces: Fortran, C++,
and \mma, so most programming tastes should be served.

Using \LT\ functions in Fortran and C++ is very similar. In Fortran,
it is necessary to include the two files {\tt tools.F} and {\tt tools.h},
the latter one in every function or subroutine. In C++, {\tt ctools.h}
must be included once. Before using any \LT\ function, {\tt bcaini} must
be called and at the end of the calculation {\tt bcaexi} may be called to
obtain a summary of errors. It is of course possible to change parameters
like the scale $\mu$ from dimensional regularization; this is described in
detail in the manual \cite{FCLTGuide}.

A very simple program would for instance be

\begin{center}
\begin{footnotesize}
\begin{tabular}{|l|l|l|}
\multicolumn{2}{l}{Fortran} & \multicolumn{1}{l}{C++} \\
\cline{1-1} \cline{3-3}
\begin{minipage}{.45\linewidth}
\begin{verbatim}

#include "tools.F"

    program simple
#include "tools.h"
    call bcaini
    print *, B0(1000D0, 50D0, 80D0)
    call bcaexi
    end

\end{verbatim}
\end{minipage} &&
\begin{minipage}{.45\linewidth}
\begin{verbatim}

#include "ctools.h"

main()
{
  bcaini();
  cout << B0(1000., 50., 80.) << "\n";
  bcaexi();
}

\end{verbatim}
\end{minipage} \\ \cline{1-1} \cline{3-3}
\end{tabular}
\end{footnotesize}
\end{center}
The \mma\ interface is even simpler to use:
\begin{verbatim}
In[1]:= Install["bca"]

In[2]:= B0[1000, 50, 80]

Out[2]= -4.40593 + 2.70414 I
\end{verbatim}

\section*{Acknowledgements}

This work has been supported by DFG under contract number Ku 502/8--1.

\begin{flushleft}

\end{flushleft}

\end{document}